\documentstyle[12pt,a4]{article}

\begin{document}

\newcommand{\nc}[2]{\newcommand{#1}{#2}}
\newcommand{\ncx}[3]{\newcommand{#1}[#2]{#3}}
\ncx{\pr}{1}{#1^{\prime}}
\nc{\nl}{\newline}
\nc{\np}{\newpage}
\nc{\nit}{\noindent}
\nc{\be}{\begin{equation}}
\nc{\ee}{\end{equation}}
\nc{\ba}{\begin{array}}
\nc{\ea}{\end{array}}
\nc{\bea}{\begin{eqnarray}}
\nc{\eea}{\end{eqnarray}}
\nc{\nb}{\nonumber}
\nc{\dsp}{\displaystyle}
\nc{\bit}{\bibitem}
\nc{\ct}{\cite}
\ncx{\dd}{2}{\frac{\partial #1}{\partial #2}}
\nc{\pl}{\partial}
\nc{\dg}{\dagger}
\nc{\cH}{{\cal H}}
\nc{\cL}{{\cal L}}
\nc{\cD}{{\cal D}}
\nc{\cF}{{\cal F}}
\nc{\cG}{{\cal G}}
\nc{\cJ}{{\cal J}}
\nc{\cQ}{{\cal Q}}
\nc{\tB}{\tilde{B}}
\nc{\tD}{\tilde{D}}
\nc{\tH}{\tilde{H}}
\nc{\tR}{\tilde{R}}
\nc{\tZ}{\tilde{Z}}
\nc{\tg}{\tilde{g}}
\nc{\tog}{\tilde{\og}}
\nc{\tGam}{\tilde{\Gam}}
\nc{\tPi}{\tilde{\Pi}}
\nc{\tcD}{\tilde{\cD}}
\nc{\tcQ}{\tilde{\cQ}}
\nc{\ag}{\alpha}
\nc{\bg}{\beta}
\nc{\gam}{\gamma}
\nc{\Gam}{\Gamma}
\nc{\bgm}{\bar{\gam}}
\nc{\del}{\delta}
\nc{\Del}{\Delta}
\nc{\eps}{\epsilon}
\nc{\ve}{\varepsilon}
\nc{\zg}{\zeta}
\nc{\th}{\theta}
\nc{\vt}{\vartheta}
\nc{\Th}{\Theta}
\nc{\kg}{\kappa}
\nc{\lb}{\lambda}
\nc{\Lb}{\Lambda}
\nc{\ps}{\psi}
\nc{\Ps}{\Psi}
\nc{\sg}{\sigma}
\nc{\spr}{\pr{\sg}}
\nc{\Sg}{\Sigma}
\nc{\rg}{\rho}
\nc{\fg}{\phi}
\nc{\Fg}{\Phi}
\nc{\vf}{\varphi}
\nc{\og}{\omega}
\nc{\Og}{\Omega}
\nc{\Kq}{\mbox{$K(\vec{q},t|\pr{\vec{q}\,},\pr{t})$ }}
\nc{\Kp}{\mbox{$K(\vec{q},t|\pr{\vec{p}\,},\pr{t})$ }}
\nc{\vq}{\mbox{$\vec{q}$}}
\nc{\qp}{\mbox{$\pr{\vec{q}\,}$}}
\nc{\vp}{\mbox{$\vec{p}$}}
\nc{\va}{\mbox{$\vec{a}$}}
\nc{\vb}{\mbox{$\vec{b}$}}
\nc{\Ztwo}{\mbox{\sf Z}_{2}}
\nc{\Tr}{\mbox{Tr}}
\nc{\lh}{\left(}
\nc{\rh}{\right)}
\nc{\ld}{\left.}
\nc{\rd}{\right.}
\nc{\nil}{\emptyset}
\nc{\bor}{\overline}
\nc{\ha}{\hat{a}}
\nc{\da}{\hat{a}^{\dg}}
\nc{\hb}{\hat{b}}
\nc{\db}{\hat{b}^{\dg}}
\nc{\hN}{\hat{N}}
\ncx{\abs}{1}{\left| #1 \right|}
\nc{\vs}{\vspace{2ex}}

\pagestyle{empty}

\begin{flushright}
NIKHEF/97-037
\end{flushright}

\begin{center}
{\LARGE {\bf Stability and mass of point particles}} \\

\vspace{4ex}

{\large J.W.\ van Holten} \\
NIKHEF-H, Amsterdam NL \\

\vspace{5ex}
september 18, 1997 \\
\vspace{5ex}

{\bf Abstract} \\
\end{center}

\nit
{\small
In this paper we consider classical point particles in full 
interaction with an arbitrary number of dynamical scalar and 
(abelian) vector fields. It is shown that the requirement of 
stability ---vanishing self-force--- is sufficient to remove 
the well-known inconsistencies of the classical theory: the 
divergent self-energy, as well as the failure of Lorentz-covariance 
of the energy-momentum when including the contributions of the 
fields. As a result, in these models the mass of a point particle 
becomes finitely computable. We discuss how these models are 
connected to quantum field theory via the path-integral 
representation of the propagator.
}

\np

\pagestyle{plain}
\pagenumbering{arabic}

\section{Introduction \label{S.1}}

The origin of particle masses is one of the recurrent themes of 
discussion in fundamental physics. The present consensus is 
that the masses of all known particles have a field-theoretical 
explanation: quark, lepton and vector boson masses are supposed 
to have their origin in the vacuum expectation value of a scalar 
field \ct{EB,Hi}. The account of the rest energy of particles is 
completed by including a contribution from the Coulomb-, Yukawa- 
and other static fields coupling to the particle. 

In the standard model, and also in the simpler case of classical 
and quantum electrodynamics, the contributions of these fields to 
the masses of particles are not computable: they are infinite, and 
infinite compensating terms have to be included in the calculations 
to get finite results for the values of the physical observables. 
These compensating terms are usually attributed to the effect of 
unknown physics at smaller distance scales. Thus particle masses 
can be accommodated in field theory, but the question whether they 
have a fully field-theoretical explanation remains open: ultimately 
the explanation of the particle spectrum is presumably to be found 
in Planck-scale physics; indeed, a truly finite theory of quantum 
gravity, e.g.\ string theory, {\em should} allow the computation of 
the mass spectrum of all particle states. Even so, in such a theory 
the masses of known particles, far below the Planck scale, might 
well have a completely field-theoretical (`low-energy') explanation. 

In this paper I explore anew the possibility of a purely 
field-theoretical explanation for (at least some) particle masses. 
I construct a class of fully interacting particle-field models in 
which the classical mass is finite and fully computable in terms 
of the self-fields of the charges carried by the particle. 
I show how mass generation (including the equivalent of the 
Brout-Engert-Higgs effect) can be incorporated in classical 
particle dynamics for the case of a particle coupled to $N_v$ 
vector fields, with vector charges $q_{\ag}$ and mass $\mu_{\ag}$ 
$(\ag = 1,...,N_v)$, and to $N_s$ scalar fields, with scalar 
charges $g_i$, mass $\mu_i$ and vacuum expectation values $f_i$ 
$(i = 1,...,N_s)$. More specifically, the following expression 
can be derived for the total particle mass, in natural units 
$(c = \hbar = 1)$: 

\be
M\, =\, m\, +\, \sum_i\, g_i f_i\, +\, \frac{1}{8\pi}\, \lh 
  \sum_i\, g_i^2 \mu_i\, -\, \sum_{\ag}\, q_{\ag}^2 \mu_{\ag} \rh, 
\label{0.1}
\ee

\nit
where $m$ is any contribution of non-field theoretical origin; 
if $m = 0$ the total mass $M$ is determined purely by the fields. 
A slightly less general form of this result (without the scalar 
vacuum expectation values) has actually been derived in the early 
days of quantum field theory \ct{pais}, but here I give a fully 
classical account: I show that the finite result hinges on the 
classical particle being stable and not subject to self-acceleration, 
thereby implying full covariance of the energy-momentum of the 
particle-field system. Thus all inconsistencies of the classical 
theory of charged particles\footnote{For a modern discussion see 
for instance ref.\ct{jack}.} are removed. 
 
The relation between this result and quantum field theory is also
discussed. An improved version of perturbation theory is outlined, 
which might preserve some of the desirable properties of the 
classical model, in particular in combination with supersymmetry. 

This paper is structured as follows. In sect.\ 2, two definitions 
of mass are recalled; it is shown how to compute them in the almost 
trivial case of a free particle. In sect.\ 3, I present a class of 
models of particles interacting with an arbitrary number of scalar 
and (abelian) vector fields. The equations of motion for a single 
particle are solved simultaneously with the field equations, 
taking full account of the back reaction of the particle acting 
as a source for the fields. It is shown that the requirement of 
stability implies two relations between the coupling constants 
and the ranges of the fields. In sect.\ 4, the energy-momentum 
tensor of the particle and its fields is computed, and it is 
shown that the stability condition implies both finiteness of 
the total mass and covariance of the total energy-momentum. 
In sect.\ 5 the mass is computed by the Hamilton-Jacobi method, 
giving the same result, eq.(\ref{0.1}). In sect.\ 6, the 
connection with quantum field theory is made using the 
path-integral formalism for the (full) propagator of the 
corresponding model. In sect.\ 7, I discuss the results 
and draw some conclusions. 
 
\section{Mass}

The equivalence principle equates the inertial an gravitational 
mass of point particles. In a special relativistic context, the 
inertial mass is defined by the kinematics, i.e.\ the dispersion 
relation between energy and momentum:

\be 
p_{\mu}^2\, +\, m^2 c^2\, =\, - \frac{E^2}{c^2}\, +\, 
 \vec{p}^{\,2}\, +\, m^2 c^2\, =\, 0. 
\label{1.0}
\ee

\nit  
The gravitational mass is defined by the energy-momentum tensor of 
the particle, acting as the source for gravitational fields in the 
Einstein equations. For a direct comparison with (\ref{1.0}), we 
should also consider it in the special relativistic limit of flat 
Minkowski space. In this limit it is a symmetric, divergence-free 
tensor field $T_{\mu\nu}(x)$: $\pl_{\mu} T^{\mu\nu} = 0$, with the 
property that, for the space-like 3-dimensional hypersurface $\Sg$: 
$x^0 =$ constant, the conserved four-momentum of eq.(\ref{1.0}) is 

\be 
p^{\mu}\, =\, \frac{1}{c}\, \int_{\Sg} d^3 x\, T^{\mu 0}.
\label{1.1}
\ee

\nit
In particular, in the rest frame $(\vec{p} = 0)$

\be
mc^2\, =\, \int_{\Sg} d^3 x\, T^{00},  
\label{1.2}
\ee

\nit
provided the integral on the r.h.s.\ of eq.(\ref{1.1}) is 
well-defined, transforming as a contravariant four-vector under 
Lorentz transformations. 

As an illustration, and as a preparation for the more complicated 
models to be considered later, I first discuss the case of the free 
point mass, described by the action \ct{AE}

\be
S_0\, =\, -\, mc^{2}\, \int d\lb \, \sqrt{- \lh \frac{1}{c} 
              \frac{d\xi^{\mu}}{d\lb} \rh^{2} }.
\label{2.10.2}
\ee

\nit
Here $\xi^{\mu}(\lb)$ are the co-ordinates of the particle as 
a function of the worldline-parameter $\lb$. Note that the 
action is actually reparametrization invariant; a natural 
and common choice for $\lb$ is to equate it to proper time:

\be 
d \lb\, =\, d\tau\, \equiv\, \frac{1}{c}\, \sqrt{- d\xi_{\mu}^2 }. 
\label{1.3}
\ee

\nit
The canonical momentum conjugate to $\xi^{\mu}$ is 

\be
p^{\mu}\, =\, m u^{\mu}\, =\, m \frac{d \xi^{\mu}}{d \tau}.
\label{2.9}
\ee

\nit
By definition of $\tau$ it satisfies the mass-shell condition 
(\ref{1.0}). The momentum can also be obtained from a 
divergence-free energy-momentum tensor as in eq.(\ref{1.1}), by 
taking  

\be
\ba{lll}
T^{\mu\nu}(x) & = & \dsp{ mc\, \int d\tau \, \frac{d \xi^{\mu}}{d \tau}  
  \frac{d \xi^{\nu}}{d \tau}\, \del^{4}\lh x - \xi(\tau) \rh } \\
  &  &  \\
  & = & \dsp{ \left[ m\,  \frac{d \xi^{\mu}}{d \tau}
        \frac{d \xi^{\nu}}{d \tau}\, \del^{3}\lh
        \frac{\vec{x} - \vec{\xi}(t)}{\sqrt{ 1 - \vec{v}^{\, 2}/c^2 }} \rh
        \right]_{\xi^{0} = ct}. }
\ea
\label{2.13}
\ee

\nit
Obviously, in the rest frame $d\xi^0 = c\, d\tau$ and eq.(\ref{1.2}) 
is satisfied. 

An alternative to this scheme is provided by the Hamilton-Jacobi 
method. The conservation of four-momentum for a free particle 
allows us to write 

\be
p^{\mu}\, =\, m\, \frac{\xi^{\mu}_f - \xi^{\mu}_i}{\tau_f - \tau_i}, 
\label{1.4} 
\ee 
 
\nit
for motion during a fixed proper-time interval $(\tau_i,\tau_f)$, 
with $(\xi_i^{\mu},\xi_f^{\mu})$ representing the initial and final 
co-ordinates of the corresponding stretch of worldline. Inserting 
the solution of the equation of motion back into the action gives 

\be
S_0^{cl}\, =\, - mc \sqrt{ - \lh \xi^{\mu}_{f} - \xi^{\mu}_{i} 
 \rh^{2}}\, =\, - mc^2 \lh \tau_f - \tau_i \rh.
\label{2.16}
\ee 

\nit
From this expression the Hamilton-Jacobi equation 

\be
p_{\mu}(\tau_{f})\, =\, \dd{S_0^{cl}}{\xi^{\mu}_{f}},
\label{2.17}
\ee

\nit
can be verified directly. Thus we observe, that the constant in front 
of the proper-time interval in the classical action defines the mass.  

One of the main results obtained below is, that for particles 
interacting with scalar and vector fields in a consistent way the 
one-particle Hamilton-Jacobi function is precisely of the form 
(\ref{2.16}), with a renormalized value of the mass parameter. This 
renormalized value then represents the physical mass, as is verified 
independently from a calculation of the stress-energy tensor. 

\section{Particles in interaction with dynamical fields}

In this section we extend the previous analysis to models of a
relativistic particles interacting with $N_s$ scalar fields
$\vf_{i}$ $(i = 1,...,N_s)$, and $N_v$ vector fields $A_{\mu}^{\ag}$ 
$(\ag = 1,...,N_v)$. We take these fields to be fully dynamical,
with (a priori arbitrary) ranges $\lb_{i,\ag} = \mu_{i,\ag}^{-1}$,
whilst the scalar fields can also have a vacuum expectation value
$\langle \vf_{i} \rangle = f_{i}$. We do not consider self-interactions 
of these fields, so our vector fields are taken to be of abelian type. 
Non-abelian interactions would require the introduction of more than 
one type of particle. Thus our model could apply to a simplified version 
of the electroweak standard model based on $U(1) \times U(1)$, in which 
a (scalar) electron couples to the photon and the $Z^{0}$, but not to 
charged vector bosons $W^{\pm}$.

With these assumptions we introduce a particle model based on the
following action

\be
\ba{lll}
S_{field} & = & \dsp{ \int d^4 x\, \left\{ - \sum_{i=1}^{N_s}\, \left[  
 \frac{1}{2}\, \lh \pl_{\mu} \vf_{i} \rh^2 + \frac{\mu_{i}^2}{2}\, 
 \lh \vf_i - f_i \rh^2
 + \rg_i\, \vf_i \right] \rd }\\
 & & \\
 & & \dsp{ \ld - \sum_{\ag = 1}^{N_v} \left[ \frac{1}{4}\,
 \lh F_{\mu\nu}^{\ag} \rh^2 + \frac{\mu_{\ag}^2}{2}\, \lh A_{\mu}^{\ag}  
 \rh^2 - \frac{1}{c}\, A_{\mu}^{\ag}\, j^{\mu}_{\ag} \right] \right\}, }
\ea
\label{3.1}
\ee

\nit
where the scalar charge densities $\rg_i$ and vector current densities 
$j^{\mu}_{\ag}$ are defined by

\[
\ba{lll}
\rg_i (x) & = & \dsp{ g_i\, \int d\lb\, \sqrt{ - \lh \frac{d\xi^{\mu}}
 {d\lb} \rh^2 }\, \del^4 \lh x - \xi(\lb) \rh }\\
 & & \\
      & = & \dsp{ g_i\, \del^3 \lh \frac{\vec{x} - \vec{\xi}(t)}{
                  \sqrt{ 1 - \vec{v}^{\,2}/c^2 }} \rh, }\\
\ea 
\]

\be
\ba{lll}
j^{\mu}_{\ag}(x) & = & \dsp{ q_{\ag} c\, \int d\lb\,
  \frac{d\xi^{\mu}}{d\lb}\, \del^4 \lh x - \xi(\lb) \rh }\\
 & & \\
  & = & \dsp{ q_{\ag}\, u^{\mu}\, \del^3 \lh
 \frac{\vec{x} - \vec{\xi}(t)}{\sqrt{ 1 - \vec{v}^{\,2}/c^2 }} \rh, }\\
\ea
\label{3.2}
\ee

\nit
where $u^{\mu}$ is the four-velocity. Note that the coupling of the 
scalar fields to the scalar charge density represents a kinetic term 
for the particle of Einstein-type, with a space-time dependent mass 
$\sum\, g_i \vf_i (x)$. It is of course possible to add a separate 
kinetic term of the type $S_{0}$, as in eq.(\ref{2.10.2}), involving 
a strictly mechanical mass. However, one can derive the above models 
from a quantum field theory through the path-integral representation 
of the propagator, as for example in \ct{Strass}-\ct{Jw2}; in that 
case the additional kinetic term is absent. 

In order to compute the contributions of the fields to the mechanical
properties of the particle, we first consider the field equations

\be
\ba{rcl}
\lh - \Box + \mu_i^2 \rh \lh \vf_i - f_i \rh & = & - \rg_i, \\
   & & \\
\left[ \lh - \Box + \mu_{\ag}^2 \rh \eta^{\mu\nu} + \pl^{\mu}\pl^{\nu} 
       \right]\, A_{\nu}^{\ag} & = & \dsp{ \frac{1}{c}\, j^{\mu}_{\ag}.}
\ea
\label{3.3}
\ee

\nit
Any solution of these equations consists of a particular solution of the 
inhomogeneous Klein-Gordon or Proca equation, for which we take the 
retarded Green's function, plus a solution of the homogeneous equation. 
In the case of a particle moving with constant velocity, the retarded 
Green's functions simplify to take the form of the usual Coulomb-Yukawa 
potentials appropriately boosted to a moving frame:

\be
\ba{lll}
\vf_i (x) & = & \dsp{ \vf_i^{free}\, +\, f_i\, -\, \frac{g_i}{4\pi}\,
                \frac{e^{-\mu_i R_{ret}}}{R_{ret}}, }\\
  & & \\
A_{\mu}^{\ag}\, & = & \dsp{ A_{\mu}^{\ag\, free}\, +\, u_{\mu}\, 
       \frac{q_{\ag}}{4\pi c}\, \frac{e^{- \mu_{\ag} R_{ret}}}{R_{ret}}. }
\ea
\label{3.4}
\ee

\nit
Here the retarded distance parameter $R_{ret} = |\vec{R}_{ret}|$ is 
obtained by boosting the relative position vector $\vec{r} = 
\vec{x} - \vec{\xi}$ in the lab frame back to the rest frame. 
Hence we get

\be
R_i^{ret} \, =\, \lh \del_{ij} - \frac{v_i v_j}{\vec{v}^{\, 2}} \rh\, 
 r_j\, +\, \frac{v_i}{\sqrt{ 1 - \vec{v}^{\, 2}/c^2 }}\, 
 \lh \frac{\vec{v} \cdot \vec{r}}{\vec{v}^{\, 2}} - t \rh.
\label{3.5}
\ee

\nit
For example, if the particle sits in the origin of its rest frame, 
which moves with velocity $v$ in the direction of the $z$-axis of 
the lab system, this reduces to

\be
\vec{R}_{ret}\, =\, \lh x, y, \frac{z - vt}{\sqrt{ 1 - v^{2}/c^2 }} \rh,
\label{3.6}
\ee

\nit
and therefore

\be
R_{ret}\, = \, \sqrt{ x^{2} + y^2 + 
  \frac{\lh z - vt \rh^2}{1 - v^{2}/c^2}},
\label{3.7}
\ee

\nit
with $(ct,x,y,z)$ the co-ordinates in the lab frame. Note also, that 
the solution of the inhomogeneous Klein-Gordon equation is shifted 
by the constant $f_i$. In line with standard terminology we refer to 
the solutions $(\vf_i^{free},A_{\mu}^{\ag\, free})$ of the homogeneous 
equations as the {\em radiation fields}, the particular solution of 
the inhomogeneous equation taking the form of the {\em Coulomb} and 
{\em Yukawa field} in the vector and scalar  case, respectively. 
The static fields always accompany the particle and contribute
to its inertial and gravitational mass.

Next we turn to the equation of motion of the particle. Varying 
$\xi^{\mu}$, and allowing for a additional mechanical mass term, 
the total action is stationary if

\be
\frac{1}{c^2}\, \frac{d}{d\tau}\, \left[ \lh mc^2 +
  \sum_{i}\, g_i \vf_{i}(\xi) \rh \, \frac{d\xi^{\mu}}{d\tau} \right]\,  
  = \, -\, \sum_i\, g_i  \pl^{\mu} \vf_{i}(\xi)\, +\,
  \sum_{\ag}\, q_{\ag} F_{\ag\, \nu}^{\mu}(\xi) \frac{d\xi^{\nu}}{d\tau}.
\label{3.8}
\ee

\nit
Now we require that in the absence of external fields the free particle, 
dressed with its Coulomb-Yukawa fields, is at rest or moves at constant 
velocity: it should not exert a net force on itself and the acceleration 
must vanish. Then 

\be
\frac{d^2 \xi^{\mu}}{d\tau^2}\, =\, 0,
\label{3.9}
\ee

\nit
with the result that 

\be
\sum_i\, g_i \pl_{\nu} \vf_{i}(\xi)\, \frac{1}{c^2}\, 
 \frac{d\xi^{\nu}}{d\tau} \frac{d\xi^{\mu}}{d\tau}\, 
 =\, - \sum_i\, g_i \pl^{\mu} \vf_{i}(\xi)\, +\,
 \sum_{\ag}\, q_{\ag} F^{\mu}_{\ag\, \nu}(\xi) \frac{d\xi^{\nu}}{d\tau}.
\label{3.10}
\ee

\nit
In the rest frame, in which all fields are static, this condition 
reduces to

\be
- \sum_i\, g_i \vec{\nabla} \vf_{i}(\xi)\, +\,
  \sum_{\ag}\, q_{\ag} \vec{E}_{\ag}(\xi)\, =\, 0,
\label{3.11}
\ee

\nit
where $\vec{E}_{\ag}$ denote the electric components of the field 
strength tensor $F^{\ag}_{\mu\nu}$, and $\xi$ is the position of the 
particle, which in the rest frame is actually the origin, according 
to our conventions. Of course, each term in eq.(\ref{3.11}) is 
singular by itself, as follows from the explicit expressions for 
the fields in eq.(\ref{3.4}) upon putting the free radiation fields 
equal to zero. However, the singular parts may now cancel between 
the scalar and vector fields, making the full sum of terms vanish. 
Explicitly, eq.(\ref{3.11}) for the fields in the rest frame becomes

\be
\lim_{R \rightarrow 0}\, \vec{\nabla}\, \lh - \sum_i\, g_i^2
        \frac{e^{-\mu_i R}}{R}\, +\, \sum_{\ag}\, q_{\ag}^2
        \frac{e^{-\mu_{\ag} R}}{R} \rh = 0.
\label{3.12}
\ee

\nit
The left-hand side is a Laurent series in $R$ with a second order 
pole and a constant term, all other terms vanishing as $R \rightarrow 
0$. The residue of the $1/R^{2}$-term, and the constant term in the 
expansion are removed if and only if the following two conditions 
are satisfied:\nl
(A) for the infinite part 

\be
\sum_i\, g_i^2\, =\, \sum_{\ag}\, q_{\ag}^2;
\label{3.13}
\ee

\nit
(B) for the finite part 

\be
\sum_i\, g_i^2 \mu_i^2\, =\, \sum_{\ag}\, q_{\ag}^2 \mu_{\ag}^2.
\label{3.14}
\ee

\nit
Therefore in these models the following observations hold: \nl
-- stability condition (A) requires both vector and scalar fields
to be present; \nl
-- if all vector fields are massless, condition (B) requires all scalar 
fields should be massless as well; \nl
-- conversely, if one or more scalar fields have a non-zero mass, (B) 
implies that the particle must couple to at least one {\em massive} vector 
field (and vice versa); for example, if our scheme would apply to 
some kind of neutrino's, the coupling of the neutrino to the $Z^0$ would 
suggest that neutrino's couple also to the Higgs fields and thus have a 
mass.

We conclude, that we have found a consistent, finite solution to the 
{\em complete} system of classical dynamical equations for the particle 
and the fields, including back reaction; consistency of this solution 
requires relations between the coupling constants and masses of the 
fields of the form (\ref{3.13}) and (\ref{3.14}). 
 
\section{The stress-energy tensor} 

The stress-energy tensor of the system of particle and fields in general
admits the following decomposition:

\be
T_{\mu\nu}\, =\, T_{\mu\nu}^{particle}\, +\, T_{\mu\nu}^{scalar} +\,
                 T_{\mu\nu}^{vector}\, +\, \Lb \eta_{\mu\nu},
\label{4.1}
\ee

\nit
where the various terms refer to the contribution of the particle, the 
scalar fields and the vector fields, and $\Lb$ is an arbitrary constant, 
which is automatically conserved and hence in principle allowed.

The stress-energy tensor is a symmetric real matrix and therefore can 
be decomposed in terms of a pseudo-orthonormal set of eigenvectors 
$n_{(\lb)}$, $\lb = 0,1,2,3$, with eigenvalues $\ag_{(\lb)}$ which in 
general are functions of the space-time point and the position of the 
particle:

\be
T^{\mu}_{\:\:\nu}\, n^{\nu}_{(\lb)}\, =\, \ag_{(\lb)}\, n^{\mu}_{(\lb)}, 
  \hspace{3em} \eta_{\mu\nu}\, n^{\mu}_{(\lb)} n^{\nu}_{(\lb^{\prime})}\, 
  =\, \eta_{\lb\lb^{\prime}}.
\label{4.2}
\ee

\nit
In our model the eigenvectors are determined completely by the geometry, 
to wit the spherical symmetry in the rest frame of the particle and the 
Lorentz boost to the lab frame; therefore the eigenvectors are actually 
the same for the various contributions to $T_{\mu\nu}$ listed above. For 
a particle moving with velocity $v$ in the $z$-direction, this universal 
basis has the form

\be
\ba{lll}
n_{(0)}^{\mu} & = & \dsp{ \lh \frac{1}{\sqrt{ 1 - v^2 /c^2 }}, 0, 0,                   
 \frac{1}{\sqrt{ 1 - v^2 /c^2 }}\, \frac{v}{c} \rh , }\\
 & & \\
n_{(1)}^{\mu} & = & \dsp{ \lh \frac{(z -vt)}{R ( 1 - v^2 /c^2)}\,
                    \frac{v}{c}, \frac{x}{R}, \frac{y}{R},
                    \frac{(z -vt)}{R ( 1 - v^2 /c^2)} \rh , }\\
 & & \\ 
n_{(2)}^{\mu} & = & \dsp{ 
     \lh \frac{ - \sqrt{x^2 + y^2}}{R \sqrt{ 1-v^2 /c^2}}\, \frac{v}{c}, 
     \frac{x}{\sqrt{x^2 + y^2}}\, \frac{(z-vt)}{R \sqrt{ 1-v^2 /c^2}}, 
     \rd }\\ 
 & & \\
 & & \hspace{4em} \dsp{ \ld \frac{y}{\sqrt{x^2 + y^2}}\, 
     \frac{(z-vt)}{R \sqrt{ 1-v^2 /c^2}}, 
     \frac{ - \sqrt{x^2 + y^2}}{R \sqrt{ 1-v^2 /c^2}} \rh , }\\ 
 & & \\
n_{(3)}^{\mu} & = & \dsp{ \lh 0, \frac{-y}{\sqrt{ x^2 + y^2 }},
                          \frac{x}{\sqrt{ x^2 + y^2 }}, 0 \rh. }\\
\ea 
\label{4.3}
\ee

\nit
In these equations $R = R_{ret}$, given by (\ref{3.7}). In the rest 
frame the expressions simplify considerably and can be written in 
spherical co-ordinates as

\be
\ba{lll}
n_{(0)} & = & \lh 1, 0, 0, 0 \rh, \\
 & & \\
n_{(1)} & = & \lh 0, \sin \th \cos \vf, \sin \th \sin \vf, 
  \cos \th \rh, \\
 & & \\
n_{(2)} & = & \lh 0, \cos \th \cos \vf, \cos \th \sin \vf, - 
  \sin \th \rh, \\
 & & \\
n_{(3)} & = & \lh 0, - \sin \vf, \cos \vf, 0 \rh. \\
\ea
\label{4.3.1}
\ee

\nit
We can now decompose the stress-energy tensor in terms of this basis 
as follows

\be
T_{\mu\nu}\, =\, \sum_{\lb}\, \ag_{(\lb)}\, n_{(\lb) \mu} n_{(\lb) \nu}.
\label{4.4}
\ee

\nit
where the eigenvalues $\ag_{(\lb)}$ are Lorentz invariant. Next we 
observe, that the time-like eigenvector is the normalized four-velocity: 
$n^{\mu}_{(0)} = u^{\mu} / c$. Therefore a consistent one-particle 
theory should yield 

\be
p^{\mu}\, =\, \frac{1}{c}\, \int_{\Sg} d^3 x\, T^{\mu 0}\, =\,
              Mc\, n_{(0)}^{\mu},
\label{4.5}
\ee

\nit
where the constant $M$ represents the physical mass of the particle, 
made up from contributions of all terms in eq.(\ref{4.1}):

\be
Mc^2\, =\, c p\cdot n_{(0)}\, 
       =\, \int_{\Sg} d^3 x\, \ag_{(0)}\, n_{(0)}^0.
\label{4.6}
\ee

\nit
In particular, in the rest frame 

\be
Mc^2\, =\, \int_{\Sg} d^3 x\, \ag_{(0)}.
\label{4.6.1}
\ee

\nit
To obtain the results (\ref{4.5})--(\ref{4.6.1}) we require that the
integrals over the stress components $(\ag_{(1)},\ag_{(2)},\ag_{(3)})$ 
in the decomposition (\ref{4.4}) of $T_{\mu\nu}$ vanish. It turns out 
that this is guaranteed if condition (A), eq.(\ref{3.13}), for the 
coupling constants is satisfied. In particular, this condition gets rid 
of the factors $4/3$ which appear in the original computation of the 
ratio between electromagnetic and kinematic mass because of the Poincar\'{e} 
stresses in the classical electron theory \ct{jack}. As a result, we can 
compute the physical mass $M$ directly in the rest frame, where the 
calculation is rather simple.

A remarkable result is, that from the same condition (A) it follows, 
that the physical mass $M$ is finite. This is surprising, because the 
energy contained in the Coulomb- and Yukawa-fields is infinite, and in 
this case they add up rather than subtract. What saves the model is, 
that the interaction of the particle with its own scalar field gives 
an equally singular negative contribution, cancelling the diverging 
contribution of the pure field term. Physically this can be understood 
from the attractive character of scalar forces.

We now demonstrate these results by an explicit computation. The
contribution of the particle to the stress-energy tensor is

\be
\ba{lll}
T_{\mu\nu}^{particle} & = & \dsp{ \frac{1}{c}\, \int d\tau\, 
 \lh mc^2 + \sum_i\,  g_i \vf_i(\xi) \rh\, \frac{d\xi_{\mu}}{d\tau} 
 \frac{d\xi_{\nu}}{d\tau}\, \del^{4}\lh x - \xi(\tau) \rh }\\
 & & \\
 & = & \dsp{ \lh mc^2 + \sum_i\, g_i \vf_{i}(\xi) \rh\, \del^{3} 
 \lh \frac{\vec{x} -\vec{\xi}(t)}{\sqrt{ 1 - v^2 /c^2}} \rh\,             
 n_{(0) \mu} n_{(0) \nu}. }
\ea
\label{4.7}
\ee

\nit
Thus the only non-zero eigenvalue of the particle term in the 
stress-energy tensor is $\ag_{(0)}$, which in the rest frame becomes 
the $T_{00}$ component. From eq.(\ref{3.4}) we then obtain the rather
singular explicit expression

\be
\ag_{(0)}^{part}\, =\,\lh mc^2 + \sum_i\, g_i f_i - \sum_i\, g_i^2
        \frac{e^{- \mu_i R}}{4\pi R} \rh \del^{3}\lh \vec{R} \rh .
\label{4.7.1}
\ee

\nit
Next we consider the scalar fields. The contribution of the scalar 
fields to the stress-energy tensor takes the form

\be
T_{\mu\nu}^{scalar}\, =\, \sum_i\, \left( \pl_{\mu} \vf_i \pl_{\nu} \vf_i
 \, -\, \frac{1}{2}\, \eta_{\mu\nu}\, \left[ ( \pl_{\kg} \vf_i )^2 + 
 \mu_i^2 ( \vf_i - f_i )^2 \right] \right).
\label{4.8}
\ee

\nit
If we substitute the solution (\ref{3.4}) with the radiation field 
$\vf^{free} = 0$, we find $\ag^{sc}_{(0)} = - \ag^{sc}_{(2)} = - 
\ag^{sc}_{(3)}$, or

\be
T_{\mu\nu}^{scalar}\, =\, \ag^{sc}_{(0)}\, \lh n_{(0) \mu} n_{(0) \nu}                      
 - n_{(2) \mu} n_{(2) \nu} - n_{(3) \mu} n_{(3) \nu} \rh
                       + \ag^{sc}_{(1)}\, n_{(1) \mu} n_{(1) \nu},
\label{4.9}
\ee

\nit
with
\be
\ba{lll}
\ag^{sc}_{(0)} & = & \dsp{ \sum_i\, \frac{g_i^2}{32 \pi^2 R^4}\,
       e^{- 2 \mu_i R}\, \lh 1 + 2 \mu_i R + 2 \mu_i^2 R^2 \rh, }\\
 & & \\
\ag^{sc}_{(1)} & = & \dsp{ \sum_i\, \frac{g_i^2}{32 \pi^2 R^4}\,
                    e^{- 2 \mu_i R}\, \lh 1 + 2 \mu_i R \rh . }\\
\ea
\label{4.10}
\ee

\nit
Note that, as the eigenvalues $\ag_{(\lb)}$ are scalars, they may be 
evaluated in any reference frame, in particular in the rest frame.

The third contribution comes from the vector fields and is evaluated 
from

\be
T_{\mu\nu}^{vector}\, =\, \sum_{\ag}\, \lh F^{\ag}_{\mu\lb}
 F^{\ag\, \lb}_{\nu} + \mu_{\ag}^2 A_{\mu}^{\ag} A_{\nu}^{\ag} - 
 \frac{1}{2}\, \eta_{\mu\nu}\, \left[ \frac{1}{2} (F^{\ag}_{\kg\lb})^2 
 + \mu_{\ag}^2 (A_{\kg}^{\ag})^2 \right] \rh .
\label{4.11}
\ee

\nit
Using the explicit solution (\ref{3.4}) with $A_{\mu}^{\ag\, free} = 0$ 
leads to $\ag^{vec}_{(0)} = \ag^{vec}_{(2)} = \ag^{vec}_{(3)}$, hence

\be
T_{\mu\nu}^{vector}\, =\, \ag^{vec}_{(0)}\, \lh n_{(0) \mu} n_{(0) \nu} 
  + n_{(2) \mu} n_{(2) \nu} + n_{(3) \mu} n_{(3) \nu} \rh\,
  +\, \ag^{vec}_{(1)}\, n_{(1) \mu} n_{(1) \nu},
\label{4.12}
\ee

\nit
in which the co-efficients $\ag^{vec}_{(0)}, \ag^{vec}_{(1)}$ have the 
same form as in the case of the scalar fields, up to signs:

\be
\ba{lll}
\ag^{vec}_{(0)} & = & \dsp{ \sum_{\ag}\, \frac{q_{\ag}^2}{32 \pi^2 R^4}\,
     e^{- 2 \mu_{\ag} R}\, \lh 1 + 2 \mu_{\ag} R + 2 \mu_{\ag}^2 R^2 \rh , 
 }\\
 & & \\
\ag^{vec}_{(1)} & = & \dsp{ - \sum_{\ag}\, \frac{q_{\ag}^2}{32 \pi^2 R^4}\,
     e^{- 2 \mu_{\ag} R}\, \lh 1 + 2 \mu_{\ag} R \rh . }
\ea
\label{4.13}
\ee

\nit
Finally we observe, that the constant term $\Lb \eta_{\mu\nu}$ gives an 
equal infinite contribution to the stresses and the energy, which only 
cancels if we take $\Lb = 0$. Hence we disregard this term from now on. 
Adding all contributions we can compute the integral

\be 
\Pi^{\mu\nu}\, \equiv\, \int_{\Sg} d^3 x\, T^{\mu \nu}(x)\, =\, 
 \sum_{\lb} \int d^3x\, \ag_{(\lb)}\, n_{(\lb)}^{\mu} n_{(\lb)}^{\nu}. 
\label{4.14}
\ee

\nit
As explained in eqs.(\ref{4.4})--(\ref{4.6}), if the integral is to 
describe the four-momentum of a real particle, the only non-vanishing 
contribution to the integral must come from the $\ag_{(0)}$-component 
of the stress-energy tensor. All stress components $\ag_{(i)}$, 
$i =(1,2,3)$ must cancel under the integral. We find that this happens 
if condition (A) is satisfied: $\sum g_i^2 = \sum q_{\ag}^2$, as 
required to cancel the infinite part of the self-force. Then in the 
rest frame

\be 
\Pi^{ij}\, =\, \int_{\Sg} d^3x\, T^{ij}\, =\, 0,
\label{4.15.0}
\ee

\nit 
whilst 

\be
p^{\mu}\, =\, \Pi^{\mu 0}\, =\, (Mc, 0, 0, 0),
\label{4.15}
\ee

\nit
with

\be
M\, =\, m\, +\, \frac{1}{c^2}\, \sum_i\, g_i f_i\, 
  +\, \frac{1}{8\pi c^2}\,
  \lh \sum_i\, g_i^2 \mu_i - \sum_{\ag}\, q_{\ag}^2 \mu_{\ag} \rh.
\label{4.16}
\ee

\nit 
This is the result announced in sect.\ \ref{S.1}. Because of the 
way the calculation is organized, by making the Lorentz covariant 
decomposition (\ref{4.4}) of the stress-energy tensor and defining 
the mass by the frame-independent expression (\ref{4.6}), the integral 
is guaranteed to give a Lorentz covariant result for $p_{\mu}$. 

From expression (\ref{4.16}) it follows, that in general the physical 
mass gets contributions from each of the three possible sources: \nl 
\nit
1. the mechanical mass $m$; \nl
2. the vacuum expectation value of the scalar fields $f_i$; \nl
3. the Coulomb and Yukawa self-energy. \nl
Any of these contributions can vanish for some physical reason, leaving 
the explanation of the particle mass in unknown mechanics, in scalar 
vacuum expectation values or in self-energy. Certainly, even if we suppose 
a purely dynamical (field theoretical) explanation of mass, this does 
{\em not} have to reside directly in the vacuum expectation values of 
the scalar fields; the self-energy terms would suffice in principle. 
However, it is quite reasonable to expect that the masses of the scalar 
and vector fields themselves are related to the vacuum expectation values:

\be
\mu_i\, =\, \sum_j\, A_{ij} f_j, \hspace{3em}
\mu_{\ag}\, =\, \sum_j\, B_{\ag j} f_j,
\label{4.16.1}
\ee

\nit
where the co-efficients $A_{ij}$ and $B_{\ag j}$ are functions of the 
coupling constants between the scalar and vector fields. Then all terms 
in the equation for the physical mass $M$ become proportional to the 
vacuum expectation values of the scalar fields.

Of course, the lowest-order (v.e.v.) terms are responsible for generating 
the full (classical) mass if

\be
\sum_{i}\, g_i^2 \mu_i\, -\, \sum_{\ag}\, q_{\ag}^2 \mu_{\ag}\, =\, 0.
\label{4.17}
\ee

\nit
Unlike our earlier relations (\ref{3.13}), (\ref{3.14}), there is no 
obvious physical need for such a constraint in terms of vanishing 
self-forces or related conditions. Notice however, that the three 
constraints (\ref{3.13}), (\ref{3.14}) and (\ref{4.17}) would reduce 
to a single constraint if the masses of all scalar and vector fields 
were equal:

\be
\mu_i\, =\, \mu_{\ag}, \hspace{2em} \forall (i, \ag).
\label{4.18}
\ee

\nit
In the standard model this is certainly not the case at low energies, 
although it is trivially true in the high-energy limit where all 
boson masses vanish. But note, that relation (\ref{4.18}) is 
characteristic for supersymmetric theories, especially $N \geq 2$ 
Yang-Mills models, where the vector and scalar masses are equal as 
long as supersymmetry is unbroken. Indeed, we can interpret the result 
(\ref{4.17}) as a classical non-renormalization theorem. 

\section{Hamilton-Jacobi formulation}

In this section an alternative derivation of the mass formula 
(\ref{4.16}) is presented, based on the Hamilton-Jacobi formalism. 
As a starting point, we perform a partial integration in the 
action (\ref{3.1})

\be
\ba{lll}
S_{field} & = & \dsp{ \int d^4 x\, \left\{ - \sum_{i=1}^N\, \left[ 
 \frac{1}{2}\, \lh \vf_{i} - f_i \rh \lh -\Box + \mu_i^2 \rh 
 \lh \vf_i - f_i \rh\, +\, \rg_i\, \vf_i \right] \rd }\\
 & & \\
 & & \dsp{ \ld - \sum_{\ag = 1}^M \left[ \frac{1}{2}\, A_{\mu}^{\ag}\,
 \lh ( -\Box + \mu_{\ag}^2 ) \eta^{\mu\nu}\, +\, \pl^{\mu} \pl^{\nu} \rh
 A_{\nu}^{\ag}\, -\, \frac{1}{c}\, A_{\mu}^{\ag}\, j^{\mu}_{\ag} \right]
 \right\} } \\
 & & \\
 & & +\, \mbox{\em boundary terms}.
\ea
\label{5.1}
\ee

\nit
In the integrand we substitute the field equations (\ref{3.3}), obtaining

\be
S^{cl}_{field}\, =\, \int d^4 x\, \left\{ 
      -\, \frac{1}{2}\, \sum_i\, \left[ \rg_i f_i + \rg_i \vf_i \right]\,   
      +\, \frac{1}{2c}\, \sum_{\ag}\, j^{\mu}_{\ag} A_{\mu}^{\ag} \right\}.
\label{5.2}
\ee

\nit
Next we take the explicit solution (\ref{3.4}), with the free radiation 
fields taken to vanish, so as to describe a single non-interacting 
particle, dressed only with its Coulomb-Yukawa fields, and we use the 
expressions (\ref{3.2}) for the scalar charge and vector current 
densities. This gives

\be
\ba{lll}
S^{cl}_{field} & = & \dsp{ \int d\lb\, 
 \sqrt{ - \lh \frac{d \xi^{\mu}}{d \lb} \rh^2 }\, \times }\\
 & & \\
 & & \dsp{ \hspace{2em}
           \left[ - \sum_i\, g_i f_i\, +\, \sum_i\, \frac{g_i^2
           e^{- \mu_i R}}{8\pi R}\, -\, \sum_{\ag}\, \frac{q_{\ag}^2
           e^{- \mu_{\ag} R}}{8\pi R} \right]_{R \rightarrow 0}. }
\ea
\label{5.3}
\ee

\nit
To obtain the last line, we have substituted for the four-velocity 
$u_{\mu}$ in the vector potential the expression

\be
u_{\mu}\, =\, \frac{d\xi_{\mu}}{d\tau}\, =\, \frac{1}{\sqrt{
  -(d\xi^{\nu}/d\lb)^2}}\, \frac{d\xi_{\mu}}{d\lb}.
\label{5.4}
\ee

\nit
Taking the limit $R \rightarrow 0$ and adding the mechanical mass-term 
to the action finally gives

\be
S^{cl}\, =\, S^{cl}_{field}\, +\, S^{cl}_{0}\, =\,
           - Mc^2\, \int d\lb\, \sqrt{ - \lh \frac{d\xi^{\mu}}{d\lb} \rh^2},
\label{5.5}
\ee

\nit
with the total mass $M$ given by expression (\ref{4.16}). Note that 
in order to obtain this result it was {\em not} necessary to substitute 
the equation of motion for the particle, except that in equation 
(\ref{5.3}) we have assumed implicitly that the particle moves at 
constant velocity. Thus we may view this action as an {\em effective 
particle action} in the absence of external fields, derived by 
integrating out the fields from the full Lagrangian.

As one might expect, $S^{cl}$ is precisely of the form of the action 
for a non-interacting particle, after replacing the mechanical mass 
$m$ by the full physical mass $M$. The value of $M$ quoted above was 
derived on the assumption of constant velocity in the absence of 
external fields. Therefore, upon substitution of the solution 
of the equation of motion for a free particle, we obtain Hamilton's 
principal function 

\be
S_{0}^{cl}\, =\, - Mc \sqrt{- (\xi_{f}^{\mu} - \xi_{i}^{\mu})^2}\,
             =\, -Mc^2\, \lh \tau_f - \tau_i \rh, 
\label{5.6}
\ee

\nit
from which we derive the expression for the four-momentum

\be
p_{\mu}(\tau_f)\, =\, \dd{S^{cl}_0}{\xi^{\mu}_f}\, =\, M u_{\mu}.
\label{5.7}
\ee

\nit
This is in full agreement with our results from the analysis of the
stress-energy tensor.

\section{Quantum theory}

The models discussed so far are purely classical, and the results 
obtained may be considered as an extension and completion of the 
classical electron model of Lorentz and Abraham \ct{L,A}. In a 
quantum field-theoretical context, one would expect the results 
to be only a first approximation, with additional contributions 
coming from the quantum-polarizability of the vacuum. 

As a first step a covariant formalism is required for a quantum 
field theoretical calculation of the mass which naturally has 
the result of equation (\ref{4.16}) as its first approximation. 
In quantum field theory, particle masses appear as poles 
in the propagator. What is needed is a formalism for computing 
the value of this pole. The approach which is most close in 
spirit to the classical treatment, and is in fact a direct 
quantum-extension of the Hamilton-Jacobi procedure, is the 
path-integral formalism. In this section I describe how to 
compute various expressions for the propagator in terms of 
various forms of the classical action. It then becomes clear 
how to extract the value of the physical mass while taking 
into account strong-field effects like the contribution from 
the Coulomb and Yukawa-type fields.  

What we have learned from the Hamilton-Jacobi treatment of the 
interacting particle models is, that the classical action of the 
full theory for a single particle coupled to scalar and vector 
fields reduces to that of a free particle, with a (finitely) 
renormalized value of the mass. We expect the same for the 
case of quantum theory: the propagator of the interacting theory 
should behave like that of a free particle, with the pole shifted 
to a renormalized value of the mass. Therefore it is instructive 
to study again first the case of a free point particle of mass $m$, 
and then proceed to the interacting case. In the classical theory 
we have used a reparametrization-invariant square-root type of 
action for the particle, $S_0$ of eq.(\ref{2.10.2}). An alternative 
is provided by the quadratic action \ct{BHDV} 

\be
S_1\, =\, \frac{m}{2}\, \int d \lb\, \left[ \frac{1}{e}\, 
   \lh \frac{d \xi^{\mu}}{d \lb} \rh^{2} - c^{2} e \right].
\label{2.3}
\ee

\nit
which is also reparametrization invariant on account of including 
the einbein variable $e(\lb)$. From this the action $S_0$ can be 
derived by solving the constraint obtained by varying $S_1$ with 
respect to $e$:

\be
c^{2} e^{2} d\lb^{2}\, =\, - \lh d\xi^{\mu} \rh^{2}\, 
  \equiv\, c^{2} d\tau^{2}, 
\label{2.6}
\ee

\nit
Substitution of the two possible solutions 

\be
e\, =\, \pm\, \sqrt{ - \lh \frac{1}{c} \frac{d\xi^{\mu}}{d\lb} \rh^{2} },
\label{2.10.1}
\ee

\nit
gives the two actions 

\be
S_{\pm}\, =\, \mp\, mc^{2}\, \int d\lb \, \sqrt{- \lh \frac{1}{c} 
              \frac{d\xi^{\mu}}{d\lb} \rh^{2} }\, =\, \mp\, S_0.
\label{2.10.3}
\ee

\nit 
The two solutions are characterized by different directions of the 
world-line evolution in terms of proper time: $d\tau = \pm e d\lb$; 
therefore the actions $S_{\pm}$ can be interpreted as the action of 
a particle and an anti-particle, respectively \ct{jw5}. This follows 
not only from the reversal of the direction of the world-line, but 
also from the role of the two actions in the quantum theory, as is 
discussed next.   

We begin with the quadratic action $S_1$, eq.(\ref{2.3}), and 
establish its relation to the Feynman propagator of a free scalar 
particle: 

\be
\Del_F (x - y)\, =\, \int \frac{d^4 p}{(2\pi)^4}\,
  \frac{e^{-i p \cdot (x-y)}}{p^2 + m^2 - i\ve},
\label{6.1}
\ee

\nit
where from now on we take natural units $c = \hbar = 1$. As noted by 
Schwinger \ct{Schw}, we can write the Feynman propagator as a proper-time 
integral:

\be
\Del_F (x - y)\, =\, \frac{i}{2m}\, \int_0^{\infty} d\tau\, K(x-y|\tau),
\label{6.2}
\ee

\nit
where $K(x-y|\tau)$ is the kernel of the relativistic Schr\"{o}dinger 
equation:

\be
i \dd{}{\tau} K(x-y|\tau)\, =\,
    \frac{1}{2m}\, \lh -\Box_x + m^2 - i\ve \rh\, K(x-y|\tau),
\label{6.3}
\ee

\nit
i.e.\ the solution satisfying the initial condition

\be
\lim_{\tau \rightarrow 0} K(x-y|\tau)\, =\, \del^4 (x-y).
\label{6.4}
\ee

\nit
The explicit expression is

\be
K(x-y|\tau)\, =\, - \frac{im^2}{(2\pi \tau)^2}\,
   e^{\frac{im}{2\tau} \lh x-y \rh^2 - \frac{i}{2} \lh m - i\ve \rh \tau}.
\label{6.5}
\ee

\nit
As the kernel satisfies Huygens' principle

\be
\int d^4 \xi\, K(x-\xi|\tau_1) K(\xi-y|\tau_2)\, =\, K(x-y|\tau_1 + \tau_2),
\label{6.6}
\ee

\nit
a discretized time path-integral is obtained by re-iterating this equation
many times:

\be
K(x-y|\tau)\, =\, \int \prod_{k=1}^N d^4\xi_k\, \prod_{i=0}^N
   K(\xi_{i+1}-\xi_i|\Del \tau_i),
\label{6.6.1}
\ee

\nit
where $\xi_0 = y$, $\xi_{N+1} = x$ and $\sum_i \Del \tau_i = \tau$. 
Taking the continuum limit we arrive at a path integral expression for 
the propagator (cf.\ct{CGL}):

\be
\Del_F (x - y)\, =\, \frac{i}{2m}\, \int_0^{\infty} dT\, 
 \int_y^x D \xi^{\mu}(\tau) \, e^{\frac{i}{2} \int_0^T d\tau 
 \left\{ m\, \dot{\xi}_{\mu}^2 - m + i\ve \right\} }
\label{6.7}
\ee

\nit
The exponent is precisely the quadratic action $S_1$ after fixing the
value of the gauge degree of freedom $e(\lb) = 1$. This can be
done consistently \ct{Jw}, as the corresponding Fadeev-Popov determinant 
is just a multiplicative constant, which is removed by proper normalization.

Next we consider the Einstein action $S_+$ and inquire into its meaning in
quantum field theory. First we make an observation about its meaning at 
the classical level. Namely, this action can be considered as describing 
the motion of the particle in the laboratory frame in which $\xi^0 = ct$ is 
the time parameter, rather than a dynamical variable. This corresponds to 
the gauge choice $\lb = t$, after which the action becomes 

\be
S_+\, =\, -m\, \int dt\, \sqrt{1 - \vec{v}^{\,2}}.
\label{6.8}
\ee

\nit
In this action we can only freely vary the spatial co-ordinates $\vec{x}$.
The corresponding phase-space is spanned by these co-ordinates and the 
momenta

\be
\vec{p}\, =\, \frac{m \vec{v}}{\sqrt{1 - \vec{v}^{\,2}} }.
\label{6.9}
\ee

\nit
The time-evolution in the laboratory frame is then described by the
Hamiltonian

\be
H\, =\, \sqrt{\vec{p}^{\,2} + m^2}.
\label{6.10}
\ee

\nit
It is straightforward to check that the corresponding Hamilton 
equations correctly reproduce the laboratory-time dynamics of the 
relativistic point particle. The Hamiltonian form of the action is

\be
S_+\, =\, \int dt\, \lh \vec{p} \cdot \vec{v}
          - \sqrt{\vec{p}^{\,2} + m^2} \rh.
\label{6.11}
\ee

\nit
We assert that with $t_1 = y^0$, $t_2 = x^0$, and $\vec{v}(t) = 
d \vec{\xi}/ dt$, the Hamiltonian path integral

\be
K^+ (x-y)\, =\, \int_{\vec{y}}^{\vec{x}} D\vec{\xi}(t)\,
  \int D\vec{p}(t)\, e^{i \int_{t_1}^{t_2} dt\, \lh \vec{p} \cdot \vec{v}
  - \sqrt{\vec{p}^{\,2} + m^2} \rh},
\label{6.12}
\ee

\nit
acquires the meaning of the positive frequency part of the propagator,
whilst the action $S_-$ gives the negative frequency part, thereby
confirming our earlier interpretation of these actions in the quantum 
theory.

To prove this assertion, we first note that $K^+ (x-y)$ defined
above is a solution of the homogeneous Klein-Gordon equation, because

\be
i\dd{}{t} K^+ (x-y)\, =\, \cH_0 K^+ (x-y),
\label{6.12.1}
\ee

\nit
where $\cH_0 = \sqrt{-\Del + m^2}$. Next we recall the well-known
decomposition of the Feynman propagator into positive and negative
frequency parts

\be
\Del_F (x-y)\, =\, \th (x^0 - y^0) \Del^+ (x-y)\, -\, \th (y^0 - x^0)
                   \Del^- (x-y),
\label{6.13}
\ee

\nit
with

\be
\Del^{\pm} (x-y)\, =\, \pm \frac{i}{(2\pi)^3}\, \int \frac{d^3 p}{2\og_p}\,
 e^{\pm i \lh \vec{p} \cdot (\vec{x} - \vec{y}) - \og_p (x^0 - y^0) \rh},
\label{6.14}
\ee

\nit
where as usual $\og_p = \sqrt{\vec{p}^{\,2} + m^2}$. The positive and 
negative frequency parts satisfy the inner-product rule

\be
\Del^{\pm} (x-y)\, =\, \int d^3 \xi\, \Del^{\pm}(x - \xi)\,
\stackrel{\leftrightarrow}{\dd{}{\xi^0}}\, \Del^{\pm} (\xi - y).
\label{6.15}
\ee

\nit
Like Huygens' principle (\ref{6.6}) this equation can be reiterated
an indefinite number of times, yielding a discretized time expression
for a path integral, which in the continuum limit reduces to 
$K^+ (x-y)$ in (\ref{6.12}). Thus the path integral constructed from 
the Einstein action represents a different type of Greens function of 
the corresponding field theory than the path integral (\ref{6.7}) 
based on the quadratic action. 

The generalization of these results to particles interacting with 
a scalar and a vector field is straightforward. One looks for the 
kernel of the Schr\"{o}dinger equation

\be
i \dd{}{t} K(x-y|\tau)\, =\, \lh \hat{\cH} - i \ve \rh\, K(x-y|\tau),
\label{6.16}
\ee

\nit
where $\hat{\cH}$ is the laplacian operator in the presence of scalar 
and vector fields:

\be
\hat{\cH}\, =\, \frac{1}{2m}\, \lh - (\pl_{\mu} - q A_{\mu})^2 +
                \frac{g^2}{2}\, \vf^2 \rh.
\label{6.17}
\ee

\nit
$K(x-y|\tau)$ is to satisfy the boundary condition (\ref{6.4}) and 
the Huygens superposition principle (\ref{6.6}). The solution of 
this problem can be written as the path integral

\be
K(x-y|T)\, =\, \int D \xi^{\mu}(\tau)\, e^{\frac{i}{2} \int_0^T d\tau  
 \left\{ m \dot{\xi}_{\mu}^2 - q A \cdot \xi - \frac{g^2}{2m} \vf^2 
 - m + i\ve \right\}}.
\label{6.18}
\ee

\nit
Then the Feynman propagator for a particle in external fields in the 
interacting theory is again given by eq.(\ref{6.2}), with the integrand 
replaced by the expression (\ref{6.18}). Finally, the propagator for such 
a particle when the fields become dynamical is obtained by functional
integration over the scalar and vector fields with a density $\exp 
(iS_0^{field})$, where $S_0^{field}$ is the kinetic action of the scalar 
and vector fields.

Now consider the alternative formulation, which may be based upon the 
Hamiltonian

\be
H\, =\, \sqrt{(\vec{p} - q \vec{A})^2 + \frac{g^2}{2} \vf^2 }\, +\, q \phi,
\label{6.19}
\ee

\nit
with $\phi = A^0$. This Hamiltonian gives the same classical equations
of motion as the action in the exponent in (\ref{6.18}). However, it
is a Hamiltonian for time-evolution in the laboratory frame, rather
than proper time, and the corresponding path integral

\be
K^+ (x-y)\, =\, \int_{\vec{y}}^{\vec{x}} D\vec{\xi}(t)\, 
 \int D\vec{p}(t)\, e^{i \int_{t_1}^{t_2} dt \lh \vec{p} \cdot \vec{v}  
 - H(\vec{p}, \vec{\xi}) \rh },
\label{6.20}
\ee

\nit
is a solution of the homogeneous Klein-Gordon equation

\be
\lh \dd{}{t} - q \phi \rh^2\, K^+ (x-y)\, =\, \left[
 \lh \vec{\nabla} - q \vec{A} \rh^2 - \frac{g^2}{2} \vf^2 \right] 
 K^+ (x-y).
\label{6.21}
\ee

\nit
With dynamical scalar and vector fields, one should again perform 
a functional integral over the fields with the weight 
$\exp(iS_0^{field})$. The interesting observation, following from 
the classical Hamilton-Jacobi formalism presented above, is that 
by expanding the fields and the particle paths about the correct 
classical solutions (\ref{3.4}) and (\ref{3.9}), modulo higher order 
quantum corrections one finds that the Green's functions $\Del_F$ 
and $\Del^{\pm}$ in the interacting theory still satisfy the 
decomposition (\ref{6.13}), provided one replaces the free mass $m$ 
everywhere by the finite physical mass $M$ of eq.(\ref{4.16}). Thus 
to this approximation the light-cone structure of the theory, 
implying causality, and the invariant distinction between particles 
and anti-particles is preserved in the interacting quantum theory. 
However, further calculations to investigate higher order quantum 
corrections (loops) remain to be done.

\section{Discussion}

In this paper I have presented a consistent theory of classical 
point charges. The model is interesting in itself, because it shows 
how particle masses become computable in terms of field parameters 
(coupling constants, vacuum expectation values and characteristic 
ranges) once the particle is intrinsically stable.

At first sight, the stability criterion seems to have little 
relevance for particle physics phenomenology, even at tree level; 
however, such a comparison may be premature. First of all, we 
have chosen to analyse here the simplest model with only abelian 
couplings, because of the advantage that it can be solved completely. 
Secondly, nothing definite can be said about the scalar sector 
of the standard model: the number of scalar fields (e.g.\ Higgs 
doublets) remains unknown, and their Yukawa couplings are completely 
arbitrary (as are the masses of quarks and leptons). Also, new heavy 
gauge bosons could enter into the stability relations (\ref{3.13}), 
(\ref{3.14}). Furthermore, the effects of spin have been ignored. 
It seems likely that adding fermions to the model could further 
improve its behaviour, for example by the interplay with one or 
more supersymmetries. 

In addition, in realistic applications one has to take into account 
quantum effects, related to the many-body nature of quantum field 
theory: pair creation, (anti-)screening and renormalization. In 
general, the contributions of these effects to masses and couplings 
as computed in perturbation theory are divergent; this renders the 
classical value of the mass meaningless. Also, it is often argued 
that since only the total (effective) mass is observable, the 
contribution of scalars, vectors and vacuum expectation values 
cannot be separated and the notion of Coulomb- and Yukawa-energy 
contributing to the inertia of the particle has no operational 
meaning.

Commenting first on the latter argument, it is clear that if the 
vacuum expectation value and range of scalar and vector fields 
can change, as during phase transitions, then the relative 
contributions of fields to the stability conditions and to the 
mass vary and certainly the changes in these quantities are 
observable. At least in theory, therefore, the various contributions 
to the mass do seem to be physically distinguishable. Our results 
then imply constraints on the changes in the values of the field 
parameters during phase transitions.

As concerns the contribution of quantum effects to the mass, there 
is no a priori reason why it should {\em not} be computable, like 
the classical mass. In fact, the BPS solutions \ct{Bog,PS} in 
supersymmetric field theories are believed to provide examples of 
this. This is significant, because the stability of classical 
monopole solutions is also guaranteed precisely because of the 
interplay between vector and scalar fields \ct{tH}-\ct{Man}.
More generally, the ultra-violet divergences one encounters 
in perturbation theory are the result of short distance fields 
which cannot be controlled even if the coupling constant is 
arbitrarily small: for {\em any} non-zero value of $(g, q)$ the 
classical Yukawa/Coulomb field becomes large as soon as the 
distance approaches $R \approx g\lb_C/4\pi$, where $\lb_C = 
\hbar/Mc$ is the Compton wavelength of the particle. Therefore, 
when computing the effect of quantum fluctuations on the 
one-particle state it is obviously important to expand the fields 
around the correct classical solution, which includes the large 
short-distance Coulomb and Yukawa fields, and not about the vacuum 
state. Indeed, there is no reason to think that a naive expansion 
in weak fields close to the vacuum would be a good approximation 
to the quantum corrections at all, except for the large-distance 
part. 

Of course, even when taking into account the singular part of 
the fields in an improved perturbation theory, finiteness of the 
result for the mass is not necessarily guaranteed. In particular, 
there is an interplay with other effects, like coupling constant 
renormalization. But the remarkable properties of supersymmetric 
gauge theories involving scalars ($N \geq 2$ in four dimensions) 
may be an indication of the viability of the scheme. The necessary 
calculations certainly involve interesting physical and 
computational problems.

\end{document}